# Studio e confronto delle strutture di Apache Spark

*Valutazione nell'utilizzo dei Dataset in sostituzione agli RDD con l'ausilio di casi di studio*

Massimiliano Morrelli
Network Contacts - Via Olivetti 17 Molfetta (Ba)

## Abstract

**English**. This document is designed to study the data structures that can be used in the Apache Spark framework and to evaluate the best performing ones to implement solutions, in particular we will evaluate advantages / disadvantages deriving from the use of Dataset for job creation. The observation of the results provides further support in evaluating the use of Dataset as an alternative to RDD, in order to understand its strengths and weaknesses. The examination of the results is possible thanks to specifically designed and implemented in Java 1.8 language. The execution of the jobs, entrusted to a suitable distributed environment, will end with the comparison between execution times and results obtained.

**Italiano.** Il presente documento nasce allo scopo di studiare le strutture dati utilizzabili nel framework Apache Spark e valutare quelle più performanti per implementare soluzioni; valuteremo in particolare i vantaggi/svantaggi derivanti dall'utilizzo dei Dataset nella progettazione dei job. L'osservazione dei risultati fornisce ulteriore supporto nel valutare l'utilizzo dei Dataset in alternativa a RDD, al fine di comprederne i punti di forza e di debolezza. L'esame dei risultati è possibile in virtù di due casi appositamente pensati e implementati in linguaggio Java 1.8. L'esecuzione dei job, affidata a un adeguato ambiente distribuito, si concluderà con il confronto tra tempi di esecuzione e risultati ottenuti.

## INDICE

## 1. Premessa e scopo del documento

Il presente documento nasce allo scopo di valutare l'utilizzo dei Dataset in alternativa ai JavaRDD per capire i punti di forza e di debolezza nell'utilizzo degli stessi e in seguito valutarne l'integrazione nel progetto di ricerca BIG4MASS[1].
Tutte le implementazioni citate nel corso della dissertazione sono state realizzate in Java 1.8

## 2. Introduzione

La creazione di un Job può avvenire utilizzando le API "native" di Apache Spark, per ottimizzare i processi sia nel tempo di esecuzione, sia nella gestione delle risorse.
In particolare si è analizzato l'utilizzo dei Dataset e degli RDD.

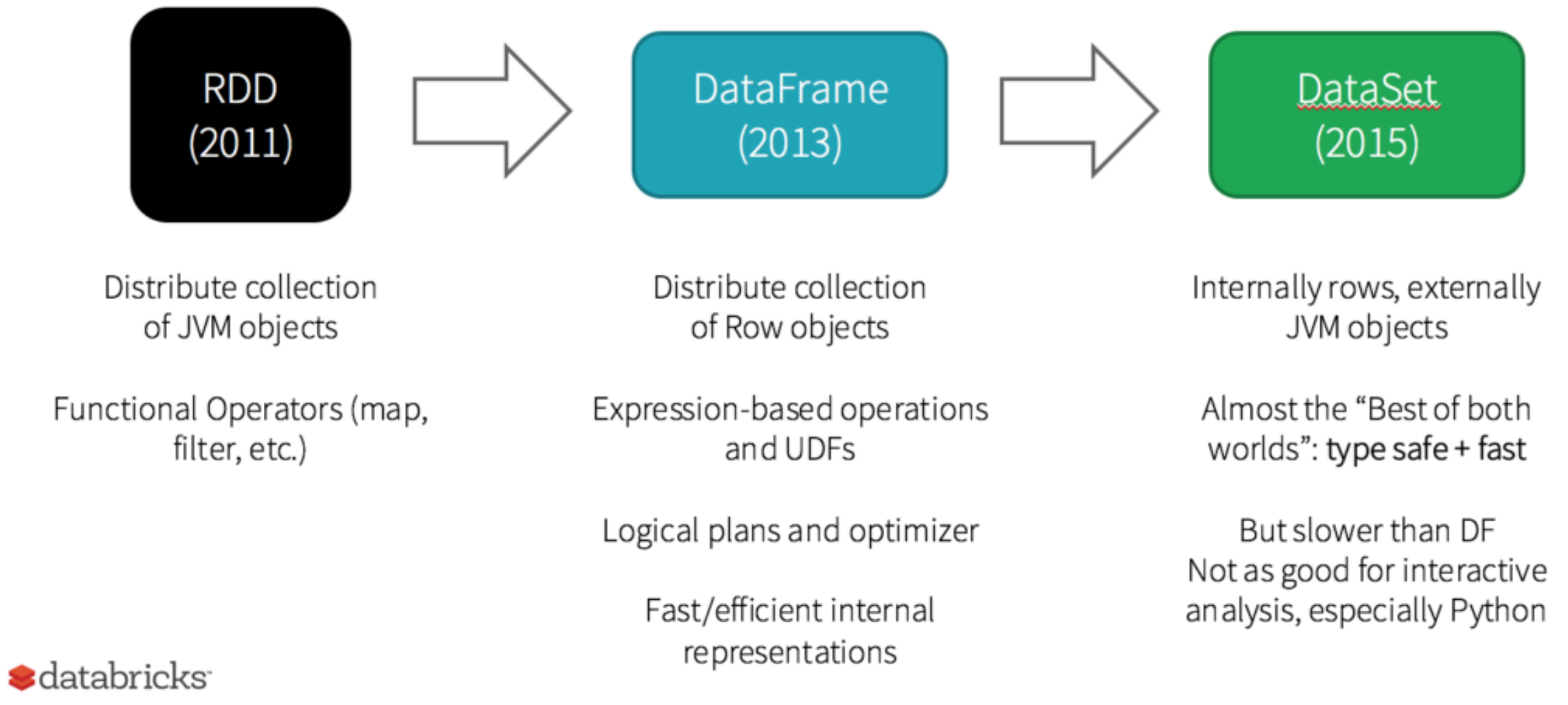

*Figura 1 - Evoluzione delle strutture in Apache Spark.*

[1] Il progetto **BIG.4.M.A.S.S. BIG** Data **F**or **M**ulti **A**gent **Specialized S**ystem, ha ad oggetto lo studio, l'analisi e la definizione di nuovi modelli di intelligenza artificiale destinati all'implementazione di sistemi ibridi e distribuiti composti da agenti virtuali e operatori umani, in grado di trattare grandi volumi di dati connotati da alta eterogenità e velocità di aggiornamento, ed a cui delegare singole attività operative/interi processi in ambito *Customer/Citizen Operation (CO).*

# 3. I tipi di API in Spark

## 3.1 RDD

Un **RDD** è una raccolta distribuita immutabile di elementi dei dati, suddivisa tra i nodi del cluster che può essere utilizzata in parallelo con un'API di basso livello che offre trasformazioni e azioni.
Un RDD (Distributed Dataset Resilient) è un'astrazione di memoria distribuita che consente ai programmatori di eseguire calcoli in memoria su cluster di grandi dimensioni in modo tollerante ai guasti.
Le caratteristiche degli RDD:

- *Resiliente* , cioè tollerante ai guasti con l'aiuto del grafico del lignaggio RDD e quindi in grado di ricalcolare le partizioni mancanti o danneggiate a causa di guasti del nodo.
- *Distribuito* con dati residenti su più nodi in un cluster .
- *Dataset* è una raccolta di dati partizionati con valori primitivi o valori di valori, ad esempio tuple o altri oggetti (che rappresentano i record dei dati con cui si lavora) (1).

Gli RDD possono eseguire due tipi di operazioni:

- le “**Trasformazioni”** che restituisce un nuovo tipo di RDD,
- le “**Azioni”** che restituiscono un valore finale al programma o scrivono dati su un sistema di archiviazione esterne.

## 3.2 Dataframe

Un ***Dataframe*** è l'API Strutturato più comune e semplicemente rappresenta una tabella di dati con righe e colonne. L'analogia più immediata è il foglio di calcolo con colonne denominate. La differenza fondamentale è che mentre un foglio di calcolo si trova su un computer in una posizione specifica, uno Spark DataFrame può estendersi su migliaia di computer. La ragione per inserire i dati su più di un computer è intuitiva: i dati sono troppo grandi per adattarsi a una macchina o la loro elaborazione richiederebbe troppo tempo su una macchina singola (2).

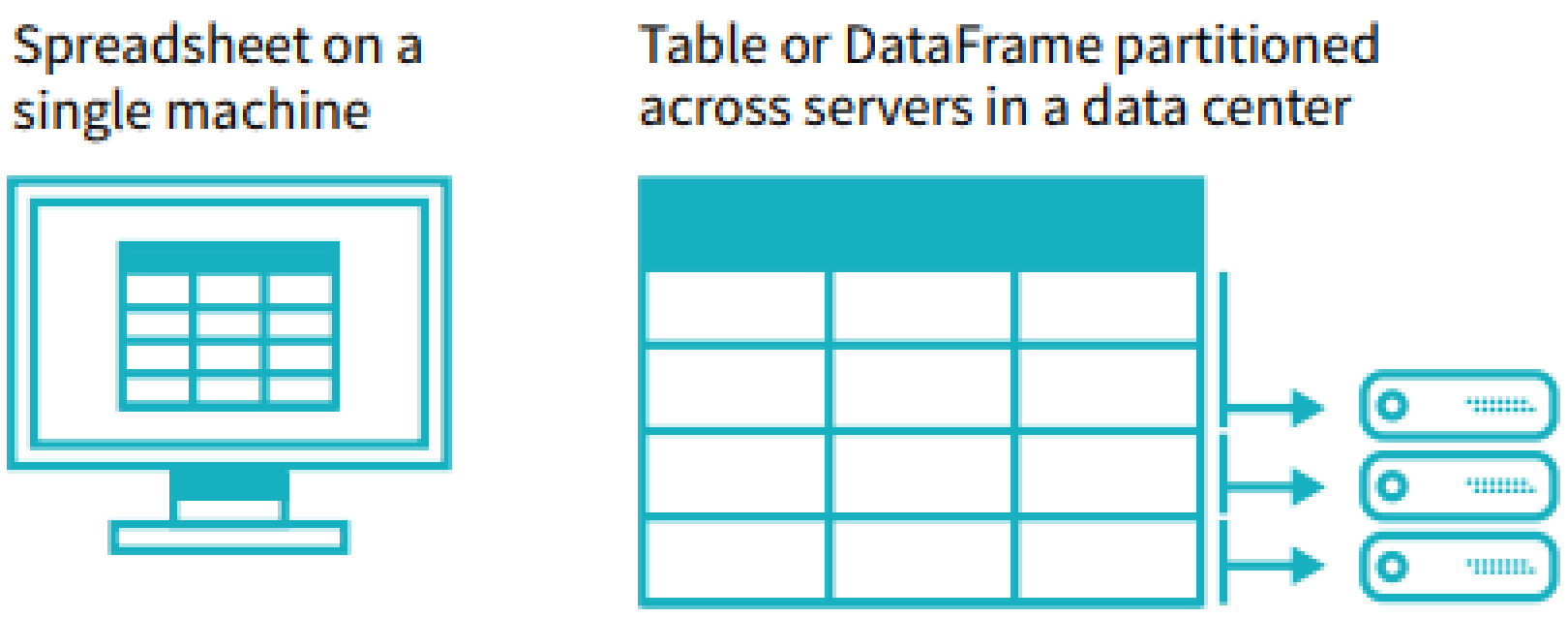

*Figura 2 - Dataframe*

Inoltre i Dataframe condividono alcune caratteristiche comuni con gli RDD:

- **Immutable**: è possibile creare un DataFrame/RDD ma non è possibile cambiarlo. È possibile trasformare un DataFrame/RDD dopo aver applicato dei metodi di trasformazione;
- **Lazy Evaluations**: un task non viene eseguito finché non viene compiuta un'azione, si tratta di una tecnica che consiste nel ritardare una computazione finché il risultato non è richiesto effettivamente;
- **Distributed**: RDD e DataFrame sono entrambi distribuiti.

## 3.3 Dataset

Un ***Dataset*** è una raccolta di oggetti fortemente tipizzata e immutabile mappata su uno schema relazionale. Il nucleo dell'API Dataset è un nuovo concetto chiamato encoder, che è responsabile della conversione tra oggetti JVM e rappresentazione tabellare. La rappresentazione tabellare viene memorizzata utilizzando il formato binario interno di Tungsten di Spark, consentendo operazioni su dati serializzati e un migliore utilizzo della memoria. Spark viene fornito con il supporto per la generazione automatica degli encoder per un'ampia varietà di tipi, compresi i tipi primitivi (ad esempio String, Integer, Long), nel caso di classi Scala e Java Beans. (3)

Un Dataset consente agli utenti di assegnare una classe Java ai record all'interno di un DataFrame e manipolarla come una raccolta di oggetti tipizzati, simile a un Java ArrayList o Scala Seq.
A partire da Spark 2.0, i tipi T supportati sono tutte le classi che seguono il modello JavaBean in Java; questi tipi sono limitati perché Spark deve essere in grado di analizzare automaticamente il tipo T e creare uno schema appropriato per i dati tabulari all'interno del set di dati (4).

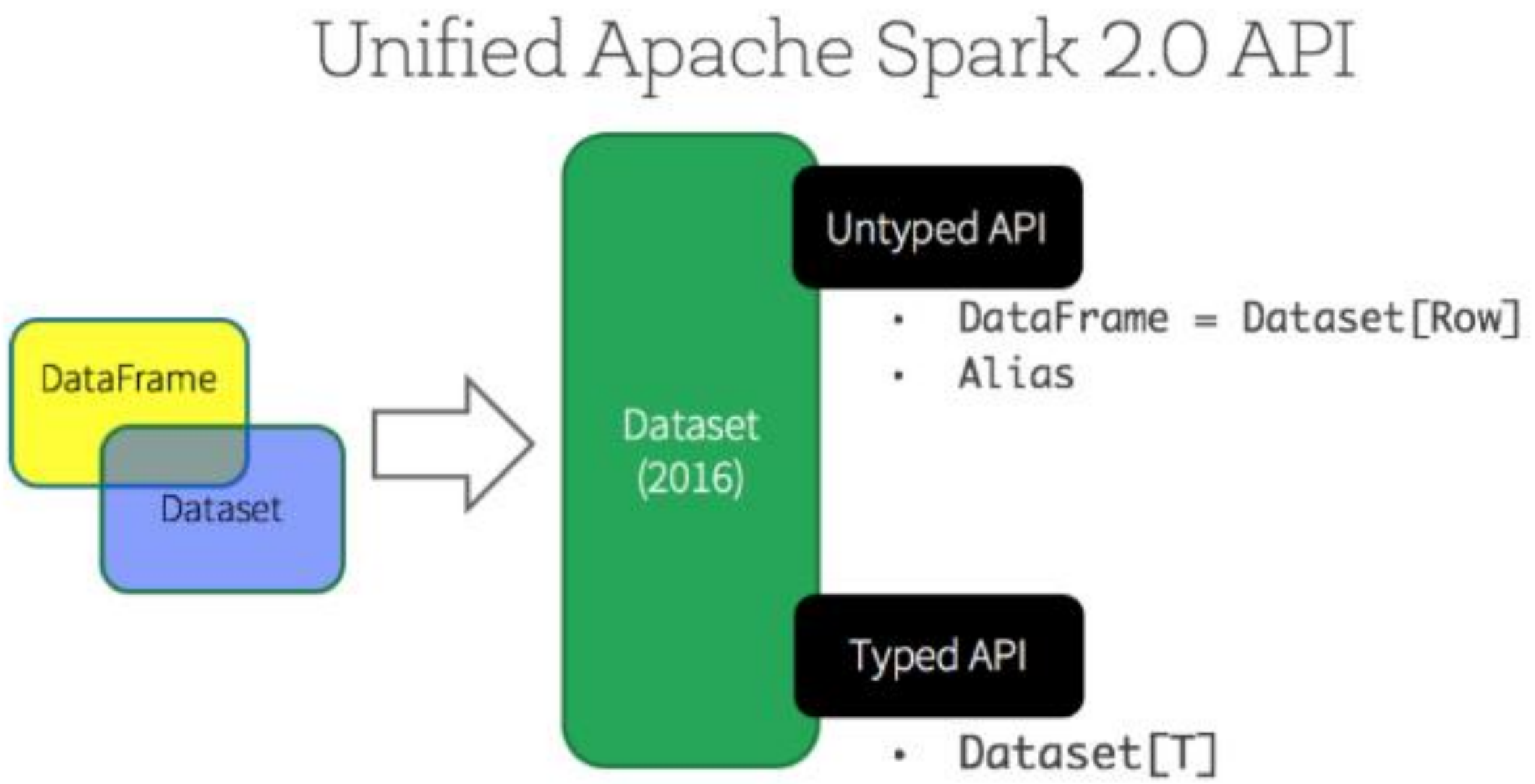

*Figura 3 - Dataset*

# 4. Casi di studio

## 4.1 Pom.xml – Dependencies

Per effettuare il confronto delle due strutture dati utilizzate, vengono implementate due soluzioni:

- NC_Spark_Reduce;
- NC_Dictionary_Json.

Di seguito vengono elencate le dipendenze utilizzate per l’implementazione delle soluzioni pensate per estrapolare i dati necessari, utili alla valutazione delle strutture di Apache Spark in osservazione.

```
<dependencies>

		<dependency>
				<groupId>org.mongodb.spark</groupId>
				<artifactId>mongo-spark-connector_2.11</artifactId>
				<version>2.3.0</version>
		</dependency>
		<dependency>
				<groupId>org.apache.spark</groupId>
				<artifactId>spark-core_2.11</artifactId>
				<version>2.3.0</version>
		</dependency>
		<dependency>
				<groupId>org.apache.spark</groupId>
				<artifactId>spark-sql_2.11</artifactId>
				<version>2.3.0</version>
		</dependency>
		<dependency>
				<groupId>org.apache.hadoop</groupId>
				<artifactId>hadoop-client</artifactId>
				<version>2.7.3</version>
		</dependency>
		<dependency>
				<groupId>junit</groupId>
				<artifactId>junit</artifactId>
				<version>4.11</version>
				<scope>test</scope>
		</dependency>
		<dependency>
				<groupId>com.networkcontacts.utils</groupId>
				<artifactId>NC_UTILS</artifactId>
				<version>0.0.2</version>
		</dependency>
```

## 4.2 NC_Spark_Reduce

NC_Spark_Reduce permette di ricavare, dato un file contenente del testo, tre collezioni (Dictionary, TwoGrams e ThreeGrams) ciascuna delle quali caratterizzata per un uso specifico.
Le collezioni possono essere generate utilizzando sia i metodi dei JavaRDD che tramite i metodi dei Dataset.
Spieghiamo brevemente la peculiarità di ciascuna collezione:

- **Dictionary** deriva dal trattamento generale del testo, dal quale si estraggono le singole parole (token) per raggrupparle in un elenco di termini distinti arricchito da due misure specifiche: la frequenza assoluta nel testo e il numero di caratteri di cui è composto il termine.
- **TwoGrams** dal testo di partenza si estrae ogni parola e l'immediata successiva, poi raggruppate in coppie distinte ciascuna con la propria frequenza assoluta.
- **ThreeGrams** dal testo di partenza si estrae ogni parola e le due che seguono, poi raggruppate in terne distinte ciascuna con la propria frequenza assoluta.

Ciascuna collezione è persistita in un database il cui nome è specificato al momento dell'esecuzione del Job.

## 4.3 NC_Dictionary_Json

Il problema è articolato intorno a un dato file JSON contenente un dizionario elettronico di grandi dimensioni che deve essere riversato sul DB.
Il file possiede diverse peculiarità:

- È molto grande e non può essere gestito con le normali applicazioni Java, si rischia facilmente di ottenere OutOfMemory;
- Non è strutturato in modo ottimale, presenta caratteristiche che rendono difficile e macchinosa l'estrazione della singola voce.

# 5. Raccolta dati

## 5.1 Raccolta dati e delle tempistiche in NC_Spark_Reduce

Un parametro utilizzato per confrontare i risultati ottenuti, sviluppando i casi di studio utilizzando sia gli RDD che i Dataset, è il tempo di svolgimento del job.

Per effettuare il confronto delle tempistiche di svolgimento del job, il team ha implementato il software denominato NC_Spark_Reduce.

Per creare le collezioni vengono analizzati due testi, il corpus[2] PAISÀ e il corpus estratto da Wikipedia, i quali difficilmente possono essere manipolati senza l'ausilio di ambienti di calcolo piuttosto potenti o sistemi distribuiti come quello che intendiamo utilizzare basato su framework Spark.

### 5.1.1 *Corpus PAISÀ*

PAISÀ è un'ampia collezione di testi autentici in lingua italiana tratti da Internet. La raccolta di testi contemporanei è stata creata nell'ambito del progetto PAISÀ (Piattaforma per l'Apprendimento dell'Italiano Su corpora Annotati) allo scopo di fornire materiale autentico e disponibile gratuitamente per l'apprendimento dell'italiano (5).

L'acquisizione delle tempistiche prodotte dalla elaborazione dei job ha prodotto i seguenti dati per il corpus di Paisa.

| Tipo | JavaRDD | Dataset | Documents |
|---|---|---|---|
| **Dictionary** | 1:58 | 1:58 | 1.762.651 |
| **TwoGrams** | 3:45 | 3:30 | 29.975.099 |
| **ThreeGrams** | 7:09 | 6:28 | 94.183.980 |

*Tabella 1 - Tempi con NC_Spark_Reduce utilizzando il corpus Paisà*

### 5.1.2 *Corpus Wikipedia*

Il corpus di Wikipedia è stato ricavato dalla omonima piattaforma on line, che consente il download del database dei contenuti in un formato specifico.

L'acquisizione delle tempistiche prodotte dalla elaborazione dei job ha prodotto i seguenti risultati.

| Tipo | JavaRDD | Dataset | Documents |
|---|---|---|---|
| **Dictionary** | 3:03 | 2.48 | 2.535.611 |
| **TwoGrams** | 6:01 | 5:16 | 45.674.085 |
| **ThreeGrams** | 12:26 | 10:40 | 150.604.408 |

*Tabella 2 - Tempi con NC_Spark_Reduce utilizzando il corpus Wikipedia*

[2] ***Corpus***: Un corpus linguistico può essere definito come una collezione di testi da utilizzare come base per analisi di natura manuale o automatica, linguistica o statistica.
Le fonti dei testi che compongono un corpus possono essere molteplici (in tal caso si parlerà di corpora misti), oppure possono coincidere con una sola tipologia testuale (corpora specialistici).

## 5.2 Raccolta dati e delle tempistiche in NC_Dictionary_Json

Il team ha realizzato due differenti implementazioni basate sui tipi JavaRDD e Dataset, entrambe elaborano opportunamente il dizionario in formato Json con tutte le difficoltà derivanti dalla struttura riccamente annidata e le tempistiche ottenute dalla elaborazione del dizionario sono:

| Tipo | JavaRDD | Dataset | Documents |
|---|---|---|---|
| **Dizionario Json** | 1:43 | 1:09 | 1.100.850 |

*Tabella 3 - Tempi con NC_Dictionary_Json*

# 6. Valutazioni sui Casi di Studio

## 6.1 Valutazioni sul utilizzo dei JavaRDD e dei Dataset

Durante lo studio dei due casi, il team, oltre al tempo di esecuzione nella creazione delle collezioni, ha valutato come parametro di osservazione anche la semplicità dell'implementazione del codice.
Nel corso della progettazione e poi stesura del codice, non si è riscontrato un netto distacco, in termini di semplicità di implementazione, ma si evidenzia che i JavaRDD sono corredati da una più ampia documentazione a sostegno dell'implementazione, mentre la documentazione per i Dataset non è così ampliamente supportata per il linguaggio Java. Infatti le varie documentazioni sono orientate a supporto di altri linguaggi di programmazione, come ad esempio Scala.
Il tempo di esecuzione di un job può essere scomposto in tre addendi:

- il *tempo di avvio* del container, $t_{container}$. Che è un tempo fisso o comunque dipendente dalle macchine che ospitano l'infrastruttura.
- il *tempo di scrittura* dei records su db di Mongo, $t_{scrittura}$,: è stato verificato in maniera empirica che non varia in base al tipo di dato utilizzato, ma dipende solo dal numero di records generati da scrivere sul db.
- Il tempo di elaborazione, $t_{elaborazione}$.

Si evince che il tempo di elaborazione, $t_{elaborazione}$, è il parametro su cui si può intervenire perché da solo subisce l'influenza del tipo di dato utilizzato.

$$t_{totale} = t_{container} + t_{elaborazione} + t_{scrittura}$$

### *6.1.1 Valutazione NC_Spark_Reduce*

Si riscontra che l'implementazione del codice nel primo caso è risultata più intuitiva con l'utilizzo dei dataset, che permettono di costruire la base dei dati in maniera pulita e lineare, senza dover manipolare le informazioni creando parametri composti da più valori da scomporre in un secondo metodo. Infatti la struttura tabellare dei dataset ha permesso di evitare alcune istruzioni cicliche e manipolazioni delle stringhe.
La creazione delle collezioni utilizzando i dataset è risultata più semplice perché le righe (row) vengono manipolate con dei metodi simili alle ListArray di Java; purtroppo quando si vuole iniziare a implementare dei dataset più complessi ci si imbatte nel concetto di Encoder[3].
Analizzando i tempi di creazione delle collezioni si evince una riduzione di tempo con l'aumentare del grado di complessità nella creazione dei documenti..
Analizzando i dati acquisiti durante l'esecuzione dei tre job si evince un risparmio maggiore per quelli che generano delle collection con molti documenti. Ad esempio, se analizziamo il progetto che ha come corpus Paisa otteniamo che la creazione della collezione di *Dictionary* avviene nello stesso tempo, mentre otteniamo un risparmio del *6,67%* nella creazione della collezione *TwoGrams* e del *9,56%* nella creazione di

[3] Encoder: Un encoder è una struttra neccessaria per definire un dataset, viene utilizzato per generare un codice in fase di runtime per serializzare l'oggetto in una struttura binaria. Questa struttura spesso ha un'impronta di memoria molto più bassa e ottimizzata per l'efficienza nell'elaborazione dei dati (11)

*ThreeGrams*, come si può evincere dal Grafico 1 - Andamento dei tempi con NC_Spark_Reduce utilizzando il corpus Paisà.

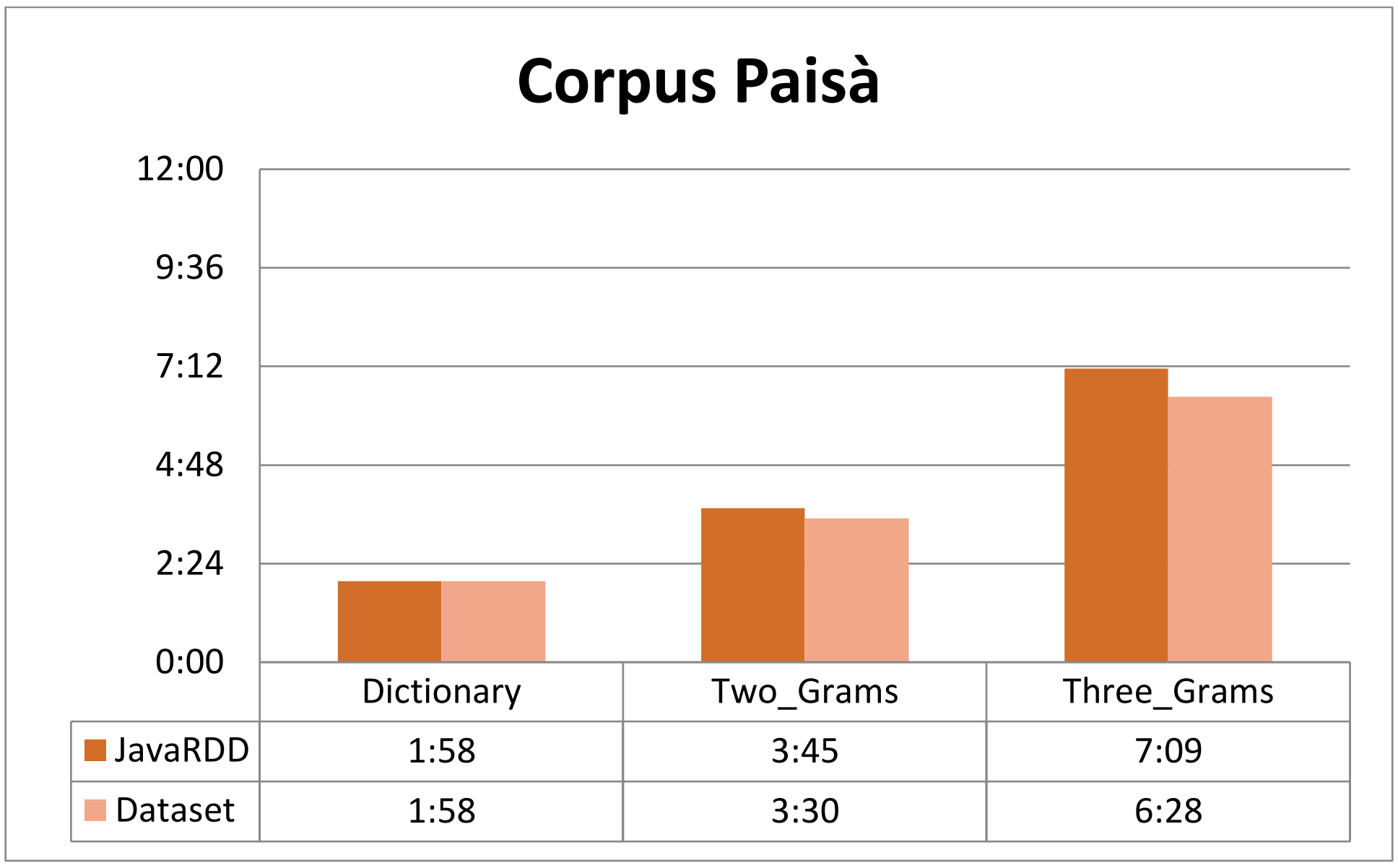

*Grafico 2 - Andamento dei tempi con NC_Spark_Reduce utilizzando il corpus Paisà*

Utilizzando come base di dati il corpus di Wikipedia si ottiene un risparmio del *8,2%* per la creazione della collezione di *Dictionary*, del *12,5%* nella creazione della collezione *TwoGrams* e infine nella produzione della collezione *ThreeGrams* si ottiene un risparmio del *14,2%* come si può evincere dal Grafico 3- Andamento dei tempi con NC_Spark_Reduce utilizzando il corpus Wikipedia.

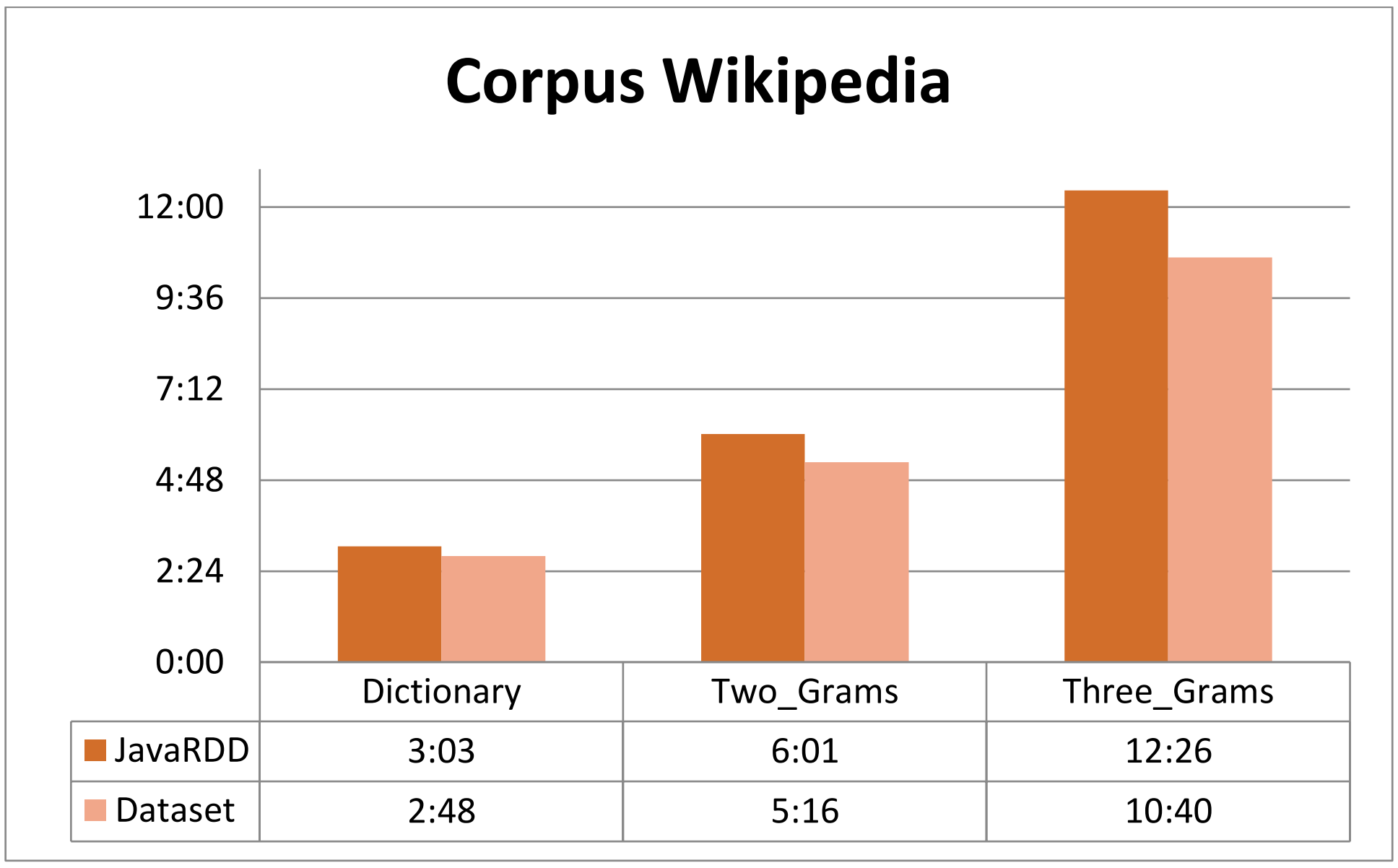

*Grafico 3- Andamento dei tempi con NC_Spark_Reduce utilizzando il corpus Wikipedia*

### *6.1.2 Valutazione NC_Dictionary_Json*

L'implementazione della soluzione NC_Dictionary_Json invece è risultata più semplice con l'ausilio dei JavaRDD rispetto ai Dataset. Ciò è dovuto dall'utilizzo di un file JSON con all'interno degli annidamenti di array che hanno aumentato la complessità nell'utilizzo dei metodi dei dataset.
Analizzando i tempi di creazione del Dizionario si evince una riduzione del tempo di analisi e della creazione della collezione. In particolare, il tempo di esecuzione si riduce del *33%*.

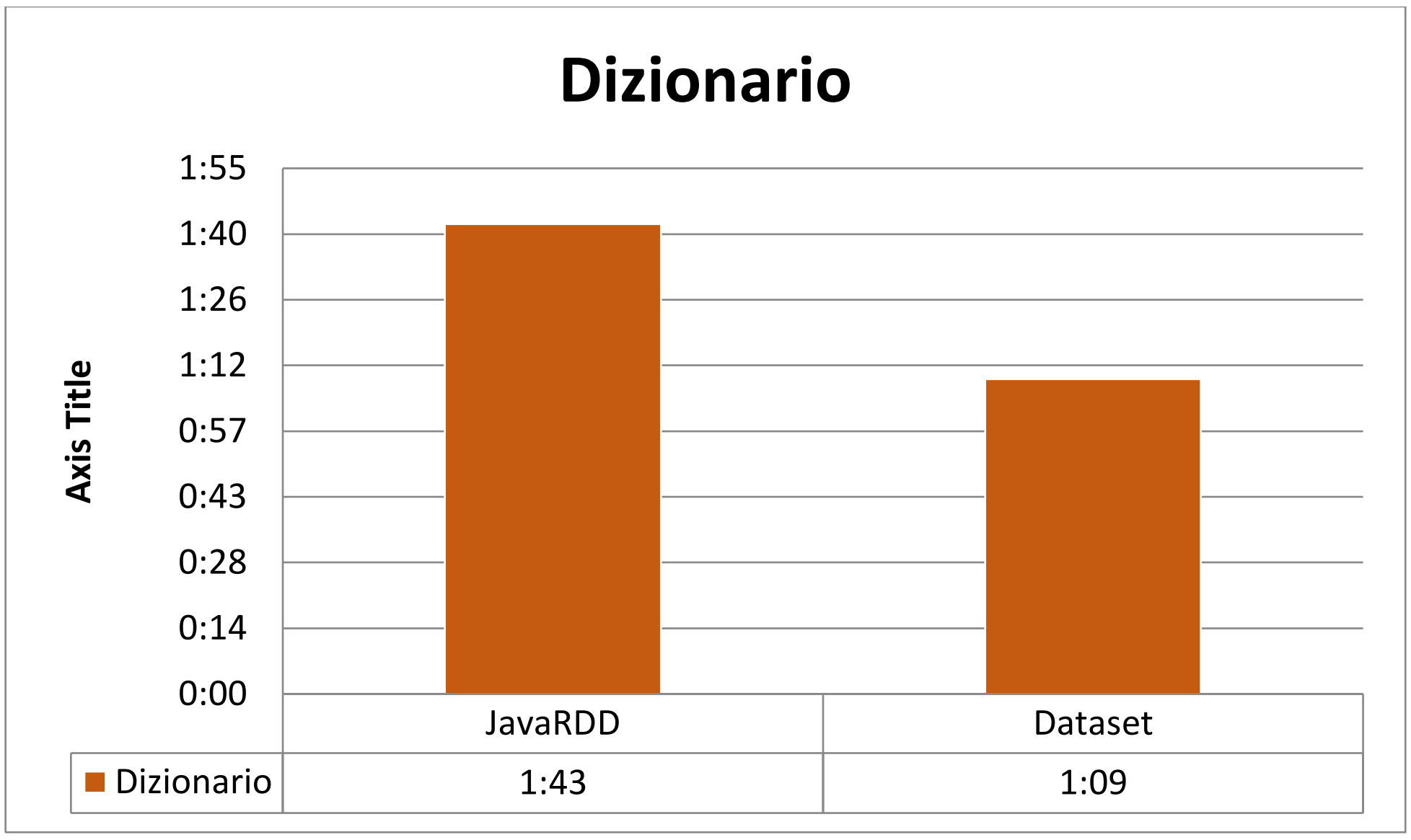

*Grafico 4 - Andamento dei tempi con NC_Dictionary_Json*

## 7. Conclusioni

In conclusione analizzando i dati ottenuti nei casi di studio, e facendo riferimento ai parametri già citati, si evince fin da subito la semplicità della struttura dei dataset che risulta molto intuitiva nel suo utilizzo.

Infatti come già spiegato nel paragrafo 3.3, i dataset si presentano come dei dati "strutturati" in forma tabellare, e ciò rende relativamente semplice la trasformazione e la manipolazione degli stessi.

Si puo notare che l'utilizzo degli encoder nell'implementazione delle soluzioni rende superfluo il codice atto a formattare i risultati, creando una struttura chiave-valore, utile al salvataggio delle collezioni su un qualsiasi database orientato ai documenti, come ad esempio MongoDB.

Di contro si evince che, i JavaRdd risultano uno strumento ideale durante l'implementazione delle soluzioni che manipolano una quantità non elevata di informazioni, impiegando un tempo di esecuzione simile a quello dei Dataset. Questa casisistica si nota nel caso della creazione del Dictionary, in NC_Spark_Reduce utilizzando il corpus di Paisà.

In questo caso l'utilizzo dei Dataset non ha fornito un valore aggiunto , anzi ha aumentato la complessità nell' implementazione della soluzione.

Si nota altresì che durante l'implementazione del codice per estrapolare il dizionario, in NC_Dictionary_Json, utilizzando i dataset ci si imbatte in strutture native di Scala, come ad esempio i WrappedArray, a differenza dell'implementazione del codice con i JavaRDD che non ha imposto l'utilizzo di altre strutture se non quelle di base.

Il presente studio integra la letteratura già esistente riguardante l'utilizzo dei Dataset e dei JavaRDD e si nota che la documentazione dei Dataset è orientata verso linguaggi di programmazione diversi da Java, come ad esempio Scala e Pyton.

La scarsa quantità di documentazione orientata alla programmazione in Java, non rende i Dataset uno strumento di facile approccio nei progetti molto complessi e strutturati.

Di conseguenza si può affermare che i Dataset sono risultati un'ottima alternativa ai JavaRDD, riducendo il tempo di elaborazione e di manipolazione di una grossa mole di informazioni, però la scelta nell'utilizzo dei Dataset o dei JavaRDD viene demandata alle priorità imposte dalla progettazione e all'uso cui è destinato il software.

## 8. Bibliografia